\documentclass[graybox]{svmult}

\usepackage{helvet}         % selects Helvetica as sans-serif font
\usepackage{courier}        % selects Courier as typewriter font
\usepackage{type1cm}        % activate if the above 3 fonts are
\usepackage{makeidx}         % allows index generation
\usepackage{graphicx}        % standard LaTeX graphics tool
\usepackage{multicol}        % used for the two-column index
\usepackage[bottom]{footmisc}% places footnotes at page bottom

\usepackage{cite}
\usepackage{amssymb}
\usepackage{amsmath}
\usepackage{amsfonts}
\usepackage{mathtools}
\usepackage{color}

\input paperdef

\begin{document}

\title*{Split NMSSM from dimensional reduction of a $10D$, $\mathcal{N}=1$, $E_8$ theory over a modified flag manifold}
\titlerunning{Split NMSSM from a $10D$, $\mathcal{N}=1$, $E_8$ theory}
\author{Gregory Patellis, Werner Porod and George Zoupanos}
\authorrunning{G. Patellis, W. Porod and G. Zoupanos}
\institute{Gregory Patellis \at Centro de Física Teórica de Partículas - CFTP, Departamento de Física, Instituto Superior Técnico, Universidade de Lisboa,
Avenida Rovisco Pais 1, 1049-001 Lisboa, Portugal, \email{grigorios.patellis@tecnico.ulisboa.pt}
\and Werner Porod \at Institut für Theoretische Physik und Astrophysik, University of Würzburg, Emil-Hilb-Weg 22, D-97074 Würzburg, Germany, \email{porod@physik.uni-wuerzburg.de}
\and George Zoupanos \at Physics Department,   National Technical University, 157 80 Zografou, Athens, Greece,\\
Max-Planck Institut f\"ur Physik, Boltzmannstr. 8, 85748 Garching, Germany,\\
Institut für Theoretische Physik der Universität Heidelberg,
Philosophenweg 16, 69120 Heidelberg, Germany, \email{george.zoupanos@cern.ch}}
\maketitle

\abstract{We review the Standard Model extension that results from the dimensional reduction of a $10D$, $\mathcal{N}=1$, $E_8$ gauge theory over the $M_4 \times SU(3)/U(1) \times U(1) \times \mathbf{Z}_3 $ space, which leads to a $4D$, $\mathcal{N}=1$,  $SU(3)^3\times U(1)^2$ theory. 
Below the unification scale we obtain a Split NMSSM effective theory.  
The third generation quark and light Higgs masses are within the experimental limits at 2-loop level and the neutralino LSP mass is predicted $<800$~GeV.}

\section{Introduction}\label{intro}
For decades, the unification of all four interactions has been a primary aim for theoretical physicists. This pursuit has been greatly emphasized by the scientific~community, leading to various intriguing approaches in recent decades. Among those, the ones with additional dimensions have received significant attention and recognition.
The idea of extra dimensions is employed by the framework of superstring theories. Within this set of theories, the heterotic string~\cite{Gross:1985fr}, which is defined in ten dimensions, stands out as particularly promising due to its potential for being experimentally tested. The phenomenological aspects of the heterotic string are very prominent in the resulting Grand Unified Theories~(GUTs), which most significantly contain the Standard Model~(SM) gauge group. These GUTs result through the compactification of the ten-dimensional spacetime and followed by a dimensional reduction on Calabi-Yau (CY) spaces.
It is important to note that an alternative~framework emerged before the formulation of superstrings, which focused on the dimensional~reduction of higher-dimensional~gauge theories, providing another path for probing the unification of fundamental~interactions. The unification of fundamental interactions was first explored by Forgacs-Manton and Scherk-Schwartz. The former developed the notion of Coset Space~Dimensional Reduction (CSDR) \cite{Forgacs:1979zs,Kapetanakis:1992hf, Kubyshin:1989vd}, while the latter worked on the group~manifold reduction \cite{Scherk:1979zr}. These were pioneering works that focused on the dimensional reduction of the extra dimensional theory and its implications on unification and set the foundation of the field.

A characteristic feature of higher-dimensional theories is that they unify gauge and scalar sectors of a $4D$ theory. Particularly in CSDR, the higher-dimensional kinetic terms of fermions result in $4D$ in addition to kinetic terms of fermions also to Yukawa terms. Furthermore, a $\mathcal{N}=1$~supersymmetric initial theory in specific numbers of dimensions can contain just one vector supermultiplet, unifying its 4D gauge and fermionic content.
Two characteristic features of the CSDR are i) that starting from a vector-like theory in $4n+2$ dimensions can lead to chiral fermions in $4D$ \cite{Manton:1981es}, ii) that $N=1$ supersymmetry in $10D$ leads to softly broken supersymmetric theories in $4D$ if the coset space used is non-symmetric and to non-supersymmetric theories if the coset is symmetric \cite{Manousselis:2001xb,Manousselis:2000aj,Manousselis:2001re,Manousselis:2004xd}.

In the heterotic string the use of compact internal CY manifolds came into prominence due to the fact that their use preserves the amount of supersymmetry of an initial $N=1$ supersymmetric gauge theory -when it is dimensionally reduced- in the remaining theory in four dimensions \cite{Candelas:1985en}.
However, their moduli~stabilization problem led to  the exploration of internal~spaces with $SU(3)$ structure, and in particular the class of nearly-Kähler~manifolds \cite{LopesCardoso:2002vpf,Strominger:1986uh,Lust:1986ix,Castellani:1986rg,Becker:2003yv,Becker:2003sh,butruille2006homogeneous}.
The $6D$ homogeneous~nearly-Kähler manifolds that admit a connection~with torsion are the non-symmetric~coset spaces $G_2/SU(3)$, $Sp(4)/SU(2)\times U(1)_{non-max}$, $SU(3)/U(1)\times U(1)$ and~the group manifold~$SU(2)\times SU(2)$~\cite{Manousselis:2005xa,Chatzistavrakidis:2008ii,Chatzistavrakidis:2009mh,Klaput:2011mz}.  
In contrast to CY, the CSDR as already mentioned can start from
a $10D$, $N=1$ theory and, using a non-symmetric coset space, can result to $4D$ theories with a naturally emergent soft supersymmetry breaking sector \cite{Manousselis:2001xb,Manousselis:2000aj,Manousselis:2001re,Manousselis:2004xd}.

In the present article we focus on the dimensional reduction of a $\mathcal{N}=1$ $E_8$ over the modified flag~manifold $SU(3)/U(1)\times U(1)\times\mathbf{Z}_3$, which is the non-symmetric coset~space $SU(3)/U(1)\times U(1)$ with the freely acting~discrete symmetry $\mathbf{Z}_3$. This choice accommodates the Wilson flux breaking~mechanism in such a way that the remaining $4D$ GUT is the $SU(3)^3\times U(1)^2$ \cite{Kapetanakis:1992hf,Manousselis:2001xb,Chatzistavrakidis:2008ii,Irges:2011de} (also \cite{Lust:1985be}).
The resulting theory is a softly~broken $\mathcal{N}=1$ theory with small radii (in a way that the compactification scale  coincides with the unification~scale). The  geometrical  origin of the soft terms renders all sfermions superheavy and they, together with the usual additional fields from the trinification group, subsequently decouple (for older versions with similar configurations see \cite{Manolakos:2020cco}, \cite{Patellis:2020cue} and \cite{Patellis:2023faw}). Due the very specific choice of radii we obtain the Split Next-to-Minimal~Supersymmetric Standard~Model (NMSSM) \cite{Demidov:2006zz,Gabelmann:2019jvz}(see also \cite{Giudice:2004tc,Ellwanger:2009dp}), in which the lighter supersymmetric~particles acquire masses $<1$TeV. For the original work see \cite{Patellis:2023npy}.

\section{The Coset Space Dimensional Reduction}\label{csdr}

We start this review with the CSDR basics. A comprehensive study of the geometric aspects of the coset spaces and the fundamental elements of CSDR (i.e. the methodology of the reduction and its constraints) can be found in \cite{Kapetanakis:1992hf,Castellani:1999fz}. Consider the $D$-dimensional space $M^4\times S/R$, where $D=d+4$ and $d$ is the number of dimensions of $S/R$. The extra dimensions of the space $M^4\times S/R$ are compactified on the coset space $S/R$, where $S$~is a Lie group~and $R$ its subgroup (and $d=dimS-dimR$). $S$ acts as a symmetry~group on the extra~coordinates. The central idea of the CSDR is the requirement that transformations of the fields under action of the symmetry group S of $S/R$ are compensated by gauge transformation. As such, since the Lagrangian considered to be gauge invariant, the result is that there is no dependence on the extra coordinates, i.e. the theory is reduced. It is worth noting that fields defined this way are called symmetric.

For a Yang-Mills-Dirac~theory  with gauge~group $G$ defined on the $D$-dimensional manifold $M^D$ and compactified~on $M^4\times S/R$, the action is
\begin{align}
S&=\int d^4xd^dy\sqrt{-g}\left[-\frac{1}{4}\text{Tr}(F_{MN}F_{K\Lambda})g^{MK}g^{N\Lambda}+\frac{i}{2}\bar{\psi} \Gamma^MD_{M}\psi\right]\,,\label{actioncsdr}
\end{align}
where $D_M$ is the covariant derivative:
\begin{equation}
    D_M=\partial_M-\theta_M-igA_M\,, \,\text{with}\,\,\, \theta_M=\frac{1}{2}\theta_{MN\Lambda}\Sigma^{N\Lambda}\,
\end{equation}
being the spin~connection and  $F_{MN}$  the field strength~tensor of~$A_M$. The spinor $\psi$ represents the fermions of the theory and belongs to the representation $F$ of the gauge group. 

The conditions that all the fields of the theory that exist on the coset space are symmetric are given as:
\begin{align}
A_{\mu}(x,y)=&g( s )A_{\mu}(x, s^{-1} y)g^{-1}( s )\nonumber\\
A_a(x,y)=&g( s )J_a^{\phantom{a}b}A_b(x, s^{-1} y)g^{-1}( s )+g( s )\partial_a g^{-1}( s )\label{constraints}\\
\psi(x,y)=&f( s )\Omega\psi(x, s^{-1} y)f^{-1}( s )\,,\nonumber\end{align}
where $g,f$ are gauge transformations in the adjoint representation $F$  of $G$, corresponding to the  $s$  transformation of  $S$  acting on  $S/R$, $J_a^{~b}$ is the  Jacobian for  $s$ and $\Omega$ is the Jacobian plus the local Lorentz rotation in tangent space. The fields $A_{\mu}$ and  $A_{\alpha}$  refer to the components of the higher dimensional gauge field $A_M = (A_{\mu}, A_{\alpha})$. The $A_{\mu}$ are the $4D$ gauge fields and the $A_{\alpha}$ are the extra-dimensional components, which "do not see" the Lorentz group, i.e. behave as $4D$ scalar fields. The above conditions imply constraints that the $D$-dimensional fields should obey.

The solutions of these constraints determine the gauge group and the surviving field content of the $4D$ theory (for details see e.g. \cite{Kapetanakis:1992hf}). The constraint referring to the $4D$ gauge fields, $A_{\mu}$ suggests that the surviving $4D$ gauge group is $H = C_G(R_G)$, i.e. the centralizer of $R$ in $G$ and $A_{\mu}$ do not depend on the coordinates of the coset.
Concerning the $4D$ scalars, $A_{\alpha}$, those that eventually survive are identified as follows:\small
\begin{equation}
G\supset R_G\times H\,,\qquad
\mathrm{adj}G=(\mathrm{adj}R,1)+(1,\mathrm{adj}H)+\sum(r_i,h_i)\,,
\end{equation}
\begin{equation}
S\supset R\,,\qquad
\mathrm{adj}S=\mathrm{adj}R+\sum s_i\,.
\end{equation}\normalsize
The scalars that survive in $4D$ are determined by the irreducible~representations $r_i$ and $s_i$ of~$R$. When $r_i$ and $s_i$ match, the representation~$h_i$ of $H$ corresponds~to a scalar~multiplet. The remaning scalars do not satisfy the constraints and are projected out i.e. do not survive in $4D$.

In a similar way, the third constraint of \eqref{constraints}, which is associated with the spinorial content of the theory, allows to determine the surviving $4D$ spinors, as the $4D$ spinors depend only on the coordinates of the $4D$ theory. The  $f_i$ representation of $H$ (to which fermions~are assigned) is determined~by the decomposition of the representation~$F$ of $G$ w.r.t. $R_G\times H$ and the spinorial representation of the local 'Lorentz group' of the tangent space, $SO(d)$, of the coset space $S/R$ under $R$ (after embedding $R$ onto $SO(d)$):\small
\begin{equation}
 G\supset R_G\times H\,,\qquad F=\sum (r_i,f_i),
\end{equation}
\begin{equation}
SO(d)\supset R\,,\qquad \sigma_d=\sum \sigma _j\,.
\end{equation}\normalsize
For each pair of identical $r_i$ and $\sigma_i$, a $f_i$ spinor  multiplet survives in $4D$.

Concerning fermions it is necessary to add few further remarks. If the higher-dimensional fermions are Dirac~fermions, the~surviving $4D$ fermions~will not be~chiral. However imposing the Weyl on an even $D$-dimensional~spacetime yields chiral fermions in $4D$. Fermions accommodated in the adjoint representation in an initial theory defined in $D=2n+2$ dimensions lead in $4D$ to two~sets of chiral fermions with identical quantum~numbers for the components of each set. If the Majorana condition is also imposed on the initial theory, the $4D$ theory does not feature the doubling of the spectrum. In order to impose both conditions (Weyl and Majorana), the initial theory has to be defined in $D=4n+2$ dimensions.

\section{Dimensional Reduction~of $E_8$ over $SU(3)/U(1)^2$}\label{E8}

We~now focus on a realistic implementation of the CSDR mechanism. We start from a $10D$, $\mathcal{N}=1$ supersymmetric, $E_8$ gauge theory  with a  vector~representation and Weyl-Majorana~fermions. According to the previous section, it is dimensionally~reduced  over the non-symmetric space~$SU(3)/U(1)\times U(1)$ \cite{Kapetanakis:1992hf,Manousselis:2001xb,Lust:1985be}. The $4D$ action after the~reduction is given:\small
\begin{align*}
    S=&C\int d^4x\,\mathrm{tr}\left[-\frac{1}{8}F_{\mu\nu}F^{\mu\nu}-\frac{1}{4}(D_\mu\phi_a)(D^\mu\phi^a)\right]+V(\phi)+\frac{i}{2}\bar{\psi}\Gamma^\mu D_\mu\psi-\frac{i}{2}\bar{\psi}\Gamma^aD_a\psi \,,
\end{align*}\normalsize
where the scalar~potential (before the constraints of \eqref{constraints} are applied) is given by
\begin{align}
    V(\phi)&=-\frac{1}{4}g^{ac}g^{bd}\mathrm{tr}\left(f_{ab}^{~~C}\phi_C-i[\phi_a,\phi_b])(f_{cd}^{~~D}\phi_D-i[\phi_c,\phi_d]\right)\label{four-dimpotential}
\end{align}
where $C$ is the coset volume, $D_\mu$ the $4D$ covariant~derivative, $D_a$ the one of the coset, the metric of the coset is $g_{\alpha\beta}=\text{diag}(R_1^2,R_1^2,R_2^2,R_2^2,R_3^2,R_3^2)$ and $R_i$ are the coset radii.

The way  $R=U(1)\times U(1)$ is embedded in $G=E_8$ determines  the $4D$ gauge~group. Our choice is that the two $U(1)$s are identified with the diagonal generators of $SU(3)$ (Cartan subalgebra) in the following maximal decomposition of  $E_8$:
\begin{align}
E_8 \supset SU(3)\times E_6 ~.
\end{align}
The gauge group in $4D$  is the centralizer of $R$ in $G$:
\begin{align}
H=C_{E_8}(U(1)_A\times U(1)_B)=E_6\times U(1)_A\times U(1)_B\,.
\end{align}
The surviving scalars and fermions are obtained by examining the 
decomposition~of the vector and~spinor representations of $SO(6)$, respectively, under~$R=U(1)_A\times U(1)_B$ (following~the methodology above).

Therefore the surviving $4D$ gauge fields fields are assigned in three $N = 1$ vector supermultiplets of the $E_6\times U(1)_A \times U(1)_B$, while the matter fields into six chiral supermultiplets, of which three are $E_6$ singlets and the other three transform under $E_6\times U(1)_A \times U(1)_B$. The unconstrained matter fields are: \small
\begin{align*}
A_i \sim 27_{(3,\frac{1}{2})}, ~~  B_i \sim
27_{(-3,\frac{1}{2})}, ~~  \Gamma_i \sim 27_{(0,-1)}, ~~
A \sim 1_{(3,\frac{1}{2})}, ~~  B \sim
1_{(-3,\frac{1}{2})}, ~~  \Gamma \sim 1_{(0,-1)}
\end{align*}\normalsize
and the scalar potential -which is positive definite- becomes:\footnotesize
\begin{align} 
V=& \frac{g^2}{2}\bigg[\frac{2}{5}\left(\frac{1}{R_1^4}+\frac{1}{R_2^4}+\frac{1}{R_3^4}\right)+\bigg(\frac{4R_1^2}{R_2^2 R_3^2}-\frac{8}{R_1^2}\bigg)\alpha^i \alpha_i + \bigg(\frac{4R_1^2}{R_2^2 R_3^2}- \frac{8}{R_1^2}\bigg)\bar{\alpha} \alpha \nonumber\\
&+\bigg(\frac{4R_2^2}{R_1^2 R_3^2}-\frac{8}{R_2^2}\bigg)\beta^i \beta_i +\bigg(\frac{4R_2^2}{R_1^2 R_3^2}-\frac{8}{R_2^2}\bigg)\bar{\beta} \beta +\bigg(\frac{4R_3^2}{R_1^2 R_2^2}-\frac{8}{R_3^2}\bigg)\gamma^i \gamma_i +\bigg(\frac{4R_3^2}{R_1^2 R_2^2}-\frac{8}{R_3^2}\bigg)\bar{\gamma} \gamma \nonumber\\
&+\bigg[80\sqrt{2} \bigg(\frac{R_1}{R_2 R_3}+\frac{R_2}{R_1 R_3}+\frac{R_3}{R_2 R_1}\bigg)d_{ijk}\alpha^i \beta^j \gamma^k+80\sqrt{2}\bigg(\frac{R_1}{R_2 R_3}+\frac{R_2}{R_1 R_3}+\frac{R_3}{R_2 R_1}\bigg)\alpha \beta \gamma +h.c \bigg] \nonumber \\
&+ \frac{1}{6}\bigg( \alpha^i(G^\alpha)_i^j\alpha_j+\beta^i(G^\alpha)_i^j\beta_j+\gamma^i(G^\alpha)_i^j\gamma_j\bigg)^2 +\frac{10}{6}\bigg( \alpha^i(3\delta_i^j)\alpha_j+\bar{\alpha}(3)\alpha+\beta^i(-3\delta_i^j)\beta_j+\bar{\beta}(-3)\beta \bigg)^2 \nonumber\\
&+\frac{40}{6}\bigg( \alpha^i(\tfrac{1}{2}\delta_i^j)\alpha_j+\bar{\alpha}(\tfrac{1}{2})\alpha+\beta^i(\tfrac{1}{2}\delta_i^j)\beta_j+\bar{\beta}(\tfrac{1}{2})\beta+\gamma^i(-1\delta_i^j)\gamma^j+\bar{\gamma}(-1)\gamma \bigg)^2 \nonumber
\end{align}
\begin{align} 
&+40\alpha^i \beta^j d_{ijk}d^{klm} \alpha_l \beta_m+40\beta^i
\gamma^j d_{ijk}d^{klm} \beta_l
\gamma_m+40 \alpha^i \gamma^jd_{ijk} d^{klm} \alpha_l \gamma_m \nonumber\\
&+40(\bar{\alpha}\bar{\beta})(\alpha\beta)+40(\bar{\beta}\bar{\gamma})(\beta\gamma)+40(\bar{\gamma}\bar{\alpha})(\gamma
\alpha)\bigg]\,,\label{E6_pot}
\end{align}\normalsize
where $\alpha^i,\alpha,\beta^i,\beta,\gamma^i,\gamma$ are the scalar~components of $A^i,B^i,\Gamma^i$ and $A,B,\Gamma$ and $d_{ijk}$ the fully~symmetric $E_6$ invariant~tensor. One can indentify $F-, D-$ and soft supersymmetry~breaking terms in the potential of \refeq{E6_pot} The $F$-terms are identified in the last two lines and come from the superpotential:
\begin{align}
\mathcal{W}(A^i,B^j,\Gamma^k,A,B,\Gamma)=\sqrt{40}d_{ijk}A^iB^j\Gamma^k+\sqrt{40}AB\Gamma\,,
\end{align}
The $D$-terms (lines~7-9 of \refeq{E6_pot}) have their usual structure and the remaining terms of \refeq{E6_pot} (except from the first term which is constant) are the soft scalar~masses and soft trilinear terms. The gaugino mass is also of~geometrical origin, although it behaves differently than the soft masses:
\begin{equation}
    M=(1+3\tau)\frac{R_1^2+R_2^2+R_3^2}{8\sqrt{R_1^2R_2^2R_3^2}}\,,\label{gaugino}
\end{equation}
For a generic~choice the gauginos~would obtain compactification scale mass \cite{Kapetanakis:1992hf}. This is prevented by the appropriate choice of the contorsion $\tau$ (details can be found in \cite{Manousselis:2001re}). Its value is chosen to be such that the following model features a electroweak (EW) scale unified gaugino mass.

\section{Wilson Flux and the Effective Unified Theory} \label{wilson}

In the above, the $27$ multiplet that accommodates the three $E_6\times U(1)_A\times U(1)_B$ supermultiplets is insufficient~to break $E_6$ to a GUT approaching the SM gauge group. To achieve further gauge breaking, the Wilson~flux breaking mechanism is~employed \cite{Kozimirov:1989kn, Zoupanos:1987wj, Hosotani:1983xw}. We will briefly review  the basics of the Wilson flux mechanism here.

In the previous, the dimensional reduction was performed over the simply connected manifold $B_0=S/R$. However, the manifold can be multiply connected. This is achieved~by considering $B=B_0/F^{S/R}$,~where $F^{S/R}$ is a~freely-acting discrete symmetry of $B_0$. For each~element $g\in F^{S/R}$, there is a corresponding~element $U_g$ in~the $4D$ gauge group $H$, which may be viewed~as the Wilson loop:
\begin{equation}
U_g={\mathcal{P}}exp\left(-i\oint_{\gamma_g} T^a A^{~a}_M dx^M \right),
\end{equation}
where $A^{~a}_M$~are the gauge fields, $\gamma_g$ a contour representing the~element $g$ of
$F^{S/R}$, $T^a$~are the generators of the group and $\mathcal{P}$  denotes the path~ordering. In the case where the considered manifold~is simply connected, the vanishing of the field strength tensor~implies that the gauge field can be set to zero through a gauge~transformation. However, when $\gamma_g$ is chosen to~be non-contractible to a point, we have $U[\gamma]\neq 1$, and the~gauge field cannot be gauged away. This means that the~vacuum field strength does not lead to $U_g=1$.~As a result, a homomorphism of $F^{S/R}$ into~$H$ is induced~with an image $T^H$, which is the subgroup of $H$ generated by~$U_g$. Furthermore, consider a field $f(x)$ defined on $B_0$.~It is evident that $f(x)$ is equivalent to another field on~$B_0$ that satisfies $f(g(x))=f(x)$ for every $g\in F^{S/R}$.~The presence of $H$ generalizes this statement:
\begin{align}
f(g(x))=U_gf(x)\,.\label{generalisedstatement}
\end{align}
Regarding the gauge~symmetry that remains by the vacuum, in the vacuum state it~is given that $A_\mu^a=0$, and consider also a gauge~transformation by the coordinate-dependent matrix $V(x)$ of $H$.~In order to keep $A_\mu^{~a}=0$ and preserve the vacuum invariance,~the matrix $V(x)$ must be chosen to be constant. Additionally,~$f\to Vf$ is consistent with \refeq{generalisedstatement}~only if
\begin{equation}
 [V,U_g]=0\, \label{consistenconditionforwilson}    
\end{equation}
for every~$g\in F^{S/R}$.~Hence, the subgroup of $H$ that remains unbroken is the~centralizer of $T^H$ in $H$. As for the matter fields that~survive in the theory, meaning the matter fields that satisfy the~condition in \refeq{generalisedstatement}, they must be~invariant under the combination:
\begin{align*}
F^{S/R}\oplus T^H.
\end{align*}
The freely-acting~discrete symmetries, $F^{S/R}$, of $B_0=S/R$ are the center of~$S$, $Z(S)$ and $W=W_S/W_R$, where it is understood that~$W_S$ and $W_R$ are the Weyl groups of $S$ and $R$, respectively.~In this case we have
\begin{equation}
F^{S/R}=\mathbb{Z}_3  \subseteq W = S_{3},
\end{equation}
since the original coset was  $B_0=SU(3)/U(1)\times U(1)$.\\

The Wilson breaking projects  the theory in a~manner that the surviving fields are the ones~that remain invariant under  $\mathbf{Z}_3$  on their gauge and~geometric indices. In our case, the $\mathbf{Z}_3$'s non-trivial~action on the gauge indices of the fields is parameterized~by the matrix
\cite{Chatzistavrakidis:2010xi}:
\begin{equation}
    \gamma_3=\text{diag}\{\mathbf{1}_3,\omega \mathbf{1}_3, \omega^2
\mathbf{1}_3\}\,,
\end{equation}
where~$\omega=e^{i\frac{2\pi}{3}}$ is the phase~that acts on the gauge fields. 
The remaining gauge fields satisfy the condition:
\begin{equation}
[A_M,\gamma_3]=0\;\;\Rightarrow A_M=\gamma_3 A_M \gamma_3^{-1}\label{filteringgaugefields}
\end{equation}
and the new gauge symmetry is $SU(3)_c\times SU(3)_L \times SU(3)_R\times U(1)_A\times U(1)_B$. The $U(1)$s are the $R$-symmetry of the~theory, which is closely~interrelated to supersymmetry.
The matter counterpart~of \refeq{filteringgaugefields} is:
\begin{equation}
A^i=\gamma_3A^i,\;\;B^i=\omega
\gamma_3B^i,\;\;\Gamma^i=\omega^2 \gamma_3\Gamma^i\,,\qquad A= A,\;\;B=\omega B,\;\;\Gamma=\omega^2 \Gamma\,.
\end{equation}\label{wilsonprojectionmatter}
By examining the~decomposition of the $27$ representation of $E_6$ under the ${SU(3)_c\times SU(3)_L\times SU(3)_R}$ gauge group, $(1,3,\bar 3)\oplus(\bar3,1,3)\oplus(3,\bar 3,1)$, one can obtain the representations of the trinification part of the gauge group in which the above fields are accommodated. Thus, the matter content of the projected~theory is:
\begin{align*}
&A_1\equiv L\sim(\mathbf{1},\mathbf{3},\overline{\mathbf{3}})_{(3,1/2)},~~B_3\equiv q^c\sim(\overline{\mathbf{3}},\mathbf{1},\mathbf{3})_{(-3,1/2)},~~\Gamma_2\equiv Q\sim(\mathbf{3},\overline{\mathbf{3}},\mathbf{1})_{(0,-1)}\\
&A\equiv\theta\sim(\mathbf{1},\mathbf{1},\mathbf{1})_{(3,1/2)}
\end{align*}
where the former three are the remaining components of $A^i,B^i,\Gamma^i$. All together they form a $27$ representation of $E_6$, which corresponds~to the content representing one generation~in the remaining theory.
To end up with three generations, one can introduce non-trivial monopole charges~in the $U(1)$s in $R$. This leads to three copies of the aforementioned fields, subsequntly resulting in three generations \cite{Dolan:2003bj}.  The trinification multiplets $L,q^c,Q$ can be now assigned and written in the more standard way:
\begin{eqnarray*}L^{(l)}=\left(\begin{array}{ccc}
 H_d^0 & H_u^+ & \nu_L \\
 H_d^- & H_u^0 & e_L \\
 \nu^c_R & e^c_R & N
 \end{array}\right)\,,\,\,
  q^{c(l)}=\left(\begin{array}{ccc}
 d^{c1}_R & u^{c1}_R & D^{c1}_R \\
 d^{c2}_R & u^{c2}_R & D^{c2}_R \\
 d^{c3}_R & u^{c3}_R & D^{c3}_R
 \end{array}
 \right)\,,\,\, Q^{(l)}=\left(\begin{array}{ccc}
 d^1_L & d^2_L & d^3_L \\
 u^1_L & u^2_L & u^3_L \\
 D^1_L & D^2_L & D^3_L
 \end{array}\right)\,,
\end{eqnarray*}
where $l=1,2,3$ is the generation index. $q^c$ and $Q$ are quark multiplets, while $L$ containts both the lepton and the Higgs sector. The quark~multiplets also contain the vector-like down-type quarks $D^{(l)}$, which~will eventually be  $SU(2)_L$ singlets, while $L$ also~features the right-handed neutrinos $\nu_R^{c(l)}$ and the~sterile neutrino-like fields $N^{(l)}$. It is useful to note that~there are three generations of Higgs doublets. Finally,~there are three trinification singlets, $\theta^{(l)}$.\\

At this point it is useful to recall that if an effective $4D$ theory is renormalizable by power counting, then it is consistent to consider it a renormalizable theory \cite{Polchinski:1983gv}. Under this light,~we respect all the symmetries and the model structure that~are derived by the higher-dimensional theory and its dimensional~reduction. We treat, however, all the parameters of the~effective theory as  free  parameters, to the extent allowed~by symmetries. In particular, all kinetic terms and $D$-terms~of the action have the gauge coupling, $g$, as dictated by the~gauge symmetry of the model. All superpotential terms must also~share the same coupling, in order to respect supersymmetry. The~freedom given by this treatment becomes apparent in the soft sector,~where each term is allowed to its own coupling.

Taking all the~above into account, the superpotential of the $SU(3)_c\times SU(3)_L\times SU(3)_R\times U(1)_A\times U(1)_B$ effective~theory will be:
\begin{align}
\mathcal{W}^{(l)}=C^{(l)}d^{abc}L_a^{(l)}q_b^{c(l)}Q_c^{(l)}\,,\label{superpot}
\end{align}
since the~$B$ and $\Gamma$ trinfication singlets were projected out by the~Wilson flux mechanism. Using the same arguments, the soft sector of the~scalar potential is now:\small
\begin{align}
V_{\text{soft}}^{(l)}=&\left(\frac{c_{L_1}^{(l)}R_1^2}{R_2^2R_3^2}-\frac{c_{L_2}^{(l)}}{R_1^2}\right)\big<L^{(l)}|L^{(l)}\big>+\left(\frac{c_{\theta_1}^{(l)}R_1^2}{R_2^2R_3^2}-\frac{c_{\theta_1}^{(l)}}{R_1^2}\right)|\theta^{(l)}|^2\nonumber \\
&+\left(\frac{c_{q_1^c}^{(l)}R_2^2}{R_1^2R_3^2}-\frac{c_{q_2^c}^{(l)}}{R_2^2}\right)\big<q^{c(l)}|q^{c(l)}\big>
+\left(\frac{c_{Q_1}^{(l)}R_3^2}{R_1^2R_2^2}-\frac{c_{Q_1}^{(l)}}{R_3^2}\right)\big<Q^{(l)}|Q^{(l)}\big>\nonumber \\
&+\left(\frac{R_1}{R_2R_3}+\frac{R_2}{R_1R_3}+\frac{R_3}{R_1R_2}\right)(c_{\alpha}^{(l)}d^{abc}L_a^{(l)}q_b^{c(l)}Q_c^{(l)}+c_{b}^{(l)}d^{abc}L_a^{(l)}L_b^{(l)}L_c^{(l)}+h.c)\nonumber\\
=& m_{L^{(l)}}^2\big<L^{(l)}|L^{(l)}\big>+m_{q^{c(l)}}^2\big<q^{c(l)}|q^{c(l)}\big>+m_{Q^{(l)}}^2\big<Q^{(l)}|Q^{(l)}\big>+m_{\theta^{(l)}}^2|\theta^{(l)}|^2\nonumber \\ &+(\alpha^{(l)abc}L_a^{(l)}q_b^{c(l)}Q_c^{(l)}+b^{(l)abc}L_a^{(l)}L_b^{(l)}L_c^{(l)}+h.c)\,,\label{soft}
\end{align}\normalsize
where $c_i^{(l)}$ are~free parameters of $\mathcal{O}(1)$. The above equation only involves the scalar components of the denoted superfields. It is evident that all sfermions, Higgs bosons and trinification singlet scalars acquire a soft mass parameter. 
Since supersymmetry~is softly broken in the model, its closely interrelated R-symmetry~can be considered softly broken as well. Thus, we take~the liberty to add R-symmetry breaking terms $L^3$, in order to~eventually generate a superheavy B-term that allows for the~standard rotation in the Higgs sector that occurs in the Split~NMSSM (see \refse{SplitNMSSM} below). These terms break R-symmetry~completely, which means that any  remaining R-parity will also~break once the R-symmetry is broken.

\section{Choice of Radii and GUT Breaking}

In the case reviewed here, the compactification scale, $M_C$ is high, thus any Kaluza-Klein modes that occur from the dimensional reduction are~irrelevant. Moreover, we consider $M_C=M_{GUT}$; this means that the coset radii are small:
\begin{eqnarray*}
R_l\sim \frac{1}{M_{GUT}}~, ~ l=1,2,3 \,.    
\end{eqnarray*}
Since the trilinear soft terms and the soft scalar masses depend on the  geometry of the coset, \refeq{soft} suggests that  they are $\sim\mathcal{O}(M_{GUT})$.  The choice $R_2=R_3$ translates to $m_{q^{c(l)}}^2=m_{Q^{(l)}}^2$, but we employ a slightly different $R_1$. Together with appropriate selection of values for $c_{\theta_i}^{(l)}$ this leads to a cancellation among terms that dictate $m_{\theta^{(3)}}^2$ and guarantees that it is $\sim\mathcal{O}(EW)$. \\

The breaking of the $SU(3)_L\times SU(3)_R\times U(1)_A\times U(1)_B$ gauge group involves the vacuum expectation values (vevs):
\[
\langle L_s^{(1)}\rangle=\left(\begin{array}{ccc}
0 & 0&0\\
0&0&0\\
0&0&V_1
\end{array}\right),\;\;
\langle L_s^{(2)}\rangle=\left(\begin{array}{ccc}
0 & 0&0\\
0&0&0\\
V_2&0&0
\end{array}\right)~,\;\;
\langle L_s^{(3)}\rangle=\left(\begin{array}{ccc}
0 & 0&0\\
0&0&0\\
V_3&0&V_4
\end{array}\right)~,
\]
\[
\langle\theta_s^{(1)}\rangle=V_5~,\;\;
\langle\theta_s^{(2)}\rangle=V_6~, 
\]
where with the $s$ index we denote the respective scalar components of the fields.  We proceed with the above-mentioned non-minimal vev content, which will prove useful in the low-energy model and we get the breaking:
\begin{equation*}
SU(3)_c\times SU(3)_L\times SU(3)_R\times U(1)_A\times U(1)_B  \xrightarrow{V_i} SU(3)_c\times SU(2)_L\times U(1)_Y~.
\end{equation*}

\section{Missing terms}\label{missing}

At tree level, the superpotential~\eqref{superpot} lacks the bilinear terms that would serve~as $\mu$-terms in the low-energy model, which would also violate the two $U(1)$s. Since R-symmetry is broken, however, a trilinear term among the Higgs doublets and the gauge singlets  emerges radiatively:
\begin{equation}
H_u^{(l)}H_d^{(l)}\overline{\theta}^{(l)}~.\label{trilinearmu}
\end{equation}
The natural generation~diagonality of \refeq{trilinearmu} leads to a~very interesting phenomenology. Since these terms are effectively $\mu$-like terms, the Higgs~doublets of the two first generations acquire a superheavy $\mu$ term,~since $\langle \theta_s^{(1,2)}\rangle\sim \mathcal{O}(GUT)$,~while the term of the third generation survives in the low-energy model. In a similar way the lepton Yukawa terms are absent at tree level, but appear radiatively. These terms also emerge via  dim-5 operators \cite{Antoniadis:2008es}:
\begin{equation}
H_u^{(l)}H_d^{(l)}\theta^{(l)}\frac{K^{(l)}}{M}~~~,~~~L^{(l)}\overline{e}^{(l)}H_d^{(l)}\frac{K^{(l)}}{M}~,
\end{equation}
where $K^{(l)}$ can be any of the fields that acquire superheavy vevs, namely $N^{(1,3)},\nu_{R}^{(2,3)}$, $\theta^{(1,2)}$ or any combination of them (provided the generation~index is respected). 
Non-negligible~operators of even higher dimension are~also present, but do not have a qualitative~impact on the model. 
An additional $\mathbb{Z}_2$ discrete symmetry in the lepton sector of the theory protects the  model from dangerous radiative or higher-dimensional terms. The rest of the allowed terms translate to superheavy masses for all trinification~singlet fields but $\theta_f^{(3)}$, which gets a $\sim \mathcal{O}(EW)$ mass due to a~cancellation among terms.

\section{The Split NMSSM Effective Theory}\label{SplitNMSSM}

We can now  sort the particle content left under the surviving SM gauge group. The  vector-like quarks~$D^{(l)}$ and the~'sterile' fields $N^{(l)}$ and~$\nu_R^{(l)}$ become supermassive and decouple, as do the singlets $\theta^{(1,2)}$ and Higgs doublets $H_{u,d}^{(1,2)}$ (this holds for fermion and scalar components). 

As discussed in \refse{E8}, the (con)torsion value is selected so that the gauginos acquire masses of a few TeV, while all sfermions get superheavy due to the geometric origin of the soft masses. 
The third generation soft Higgs mass~parameters   $m_{L^{(3)}}^2\equiv m_{H_{u,d}}^2$ are superheavy,~while the last term of \refeq{soft} contains a soft B-like term $\sim\mathcal{O}(GUT)$:
\begin{equation}
    b^{(3)}H_u^{(3)}\cdot H_d^{(3)} \equiv b H_u\cdot H_d~.
\end{equation}
Assuming a cancellation~between $m_{H_{u,d}}^2$ and $b$ (like in \cite{Gabelmann:2019jvz}),~the Higgs doublets and the singlet field of the third~generation $H_{u,d}^{(3)}\equiv H_{u,d}$ and $\theta^{(3)}\equiv S$~are light and survive down to the electroweak scale,~having an interaction term from \refeq{trilinearmu}~(in superfield notation):
\begin{equation}
\lambda S H_u\cdot H_d~\,,
\end{equation}
where the generation~indices are dropped, since we focus on the third generation from now on. 

The scenario described above is actually the split NMSSM \cite{Demidov:2006zz,Gabelmann:2019jvz} (we adopt to the notation of \cite{Gabelmann:2019jvz}). The unification scale soft Higgs mass parameter makes the heavy Higgs scalars ($H_0,A_H,H^{\pm}$)  superheavy and they decouple. As such, the light scalar~sector contains only the light Higgs boson $h$, the scalar $S$ (which will also acquire a vev) and its CP-odd counterpart~$A$.  The higher-energy theory imposes another constraint to the Yukawa sector of the model, as the struture of its superpotential implies matching top and bottom couplings at the unification scale. It is therefore natural to expect a  large value of $tan\beta$.\\

We implemented the model in \texttt{SARAH}
\cite{Staub:2013tta} and generated a corresponding  \texttt{SPheno} code \cite{Porod:2003um,Porod:2011nf}, in order to produce~the (light) particle spectrum of the~model.
The 2-loop renormalisation group equations (RGEs) use the above-mentioned relations among couplings at the unification level as boundary conditions and they  are ran down to the EW scale. 

We also take into account~threshold corrections originating from the superheavy particles~that decouple and we allow for an extra $5\%$ uncertainty on the~boundary condition of the Yukawa couplings. For the following~analysis, we use the on-shell in case of the top~quark and the $\overline{\text{MS}}$
in case of the~bottom quark: 
\begin{equation}
m_t=(172.69\pm 0.30)~\text{GeV}~~,~~~~~ m_b(m_b)=(4.18\pm 0.03)~\text{GeV}~~,\label{topbottom}
\end{equation}
as given in~\cite{ParticleDataGroup:2022pth}. As expected, in order to~satisfy these limits we have $70<\tan\beta<80$. 
Note, that~$\beta$ is the angle between $H_u$ and $H^*_d$ which determines~the light Higgs doublet at the high scale
at which~the second doublet is integrated out.
Both,~$H$ and $S$, obtain vevs denoted by $v_H$ and $v_S$, respectively.~The combination $\mu=\lambda v_S/\sqrt{2}$ is the mass parameter~for the higgsinos like fermions and has
to be~sufficiently large to be consistent with existing LHC searches,~see discussion below.

\begin{figure}[htb!]
\sidecaption
% For example, with the graphicx style use
\includegraphics[scale=.1]{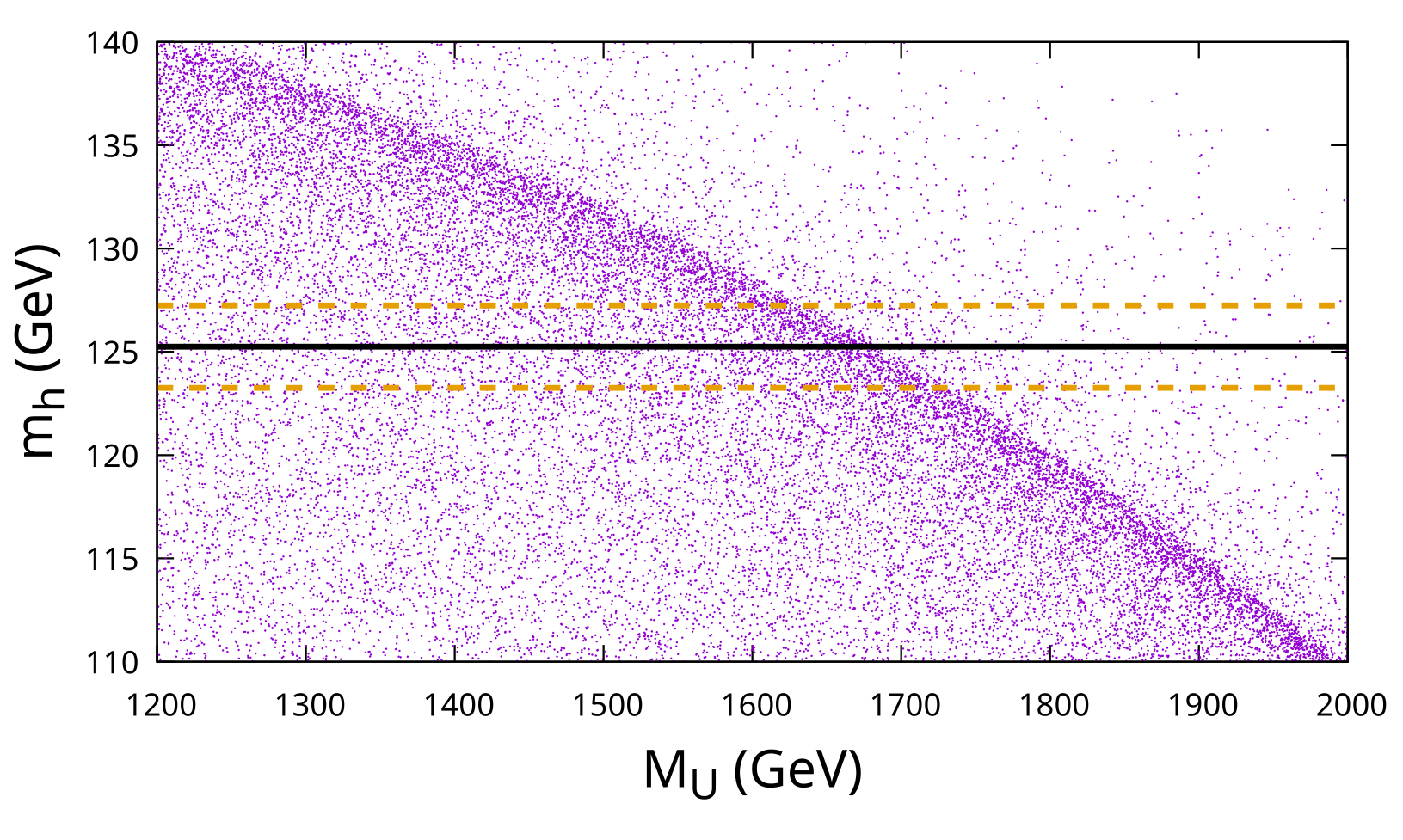}
%
% If no graphics program available, insert a blank space i.e. use
%\picplace{5cm}{2cm} % Give the correct figure height and width in cm
%
\caption{The light Higgs boson mass as a function of the unified gaugino mass. The black line denotes the experimental value of the Higgs mass, $m_h=125.25$ GeV, while the orange dashed lines denote the 2~GeV theoretical uncertainties.}
\label{fig:higgsM}       % Give a unique label
\end{figure}

\begin{figure}[htb!]
\sidecaption
% For example, with the graphicx style use
\includegraphics[scale=.1]{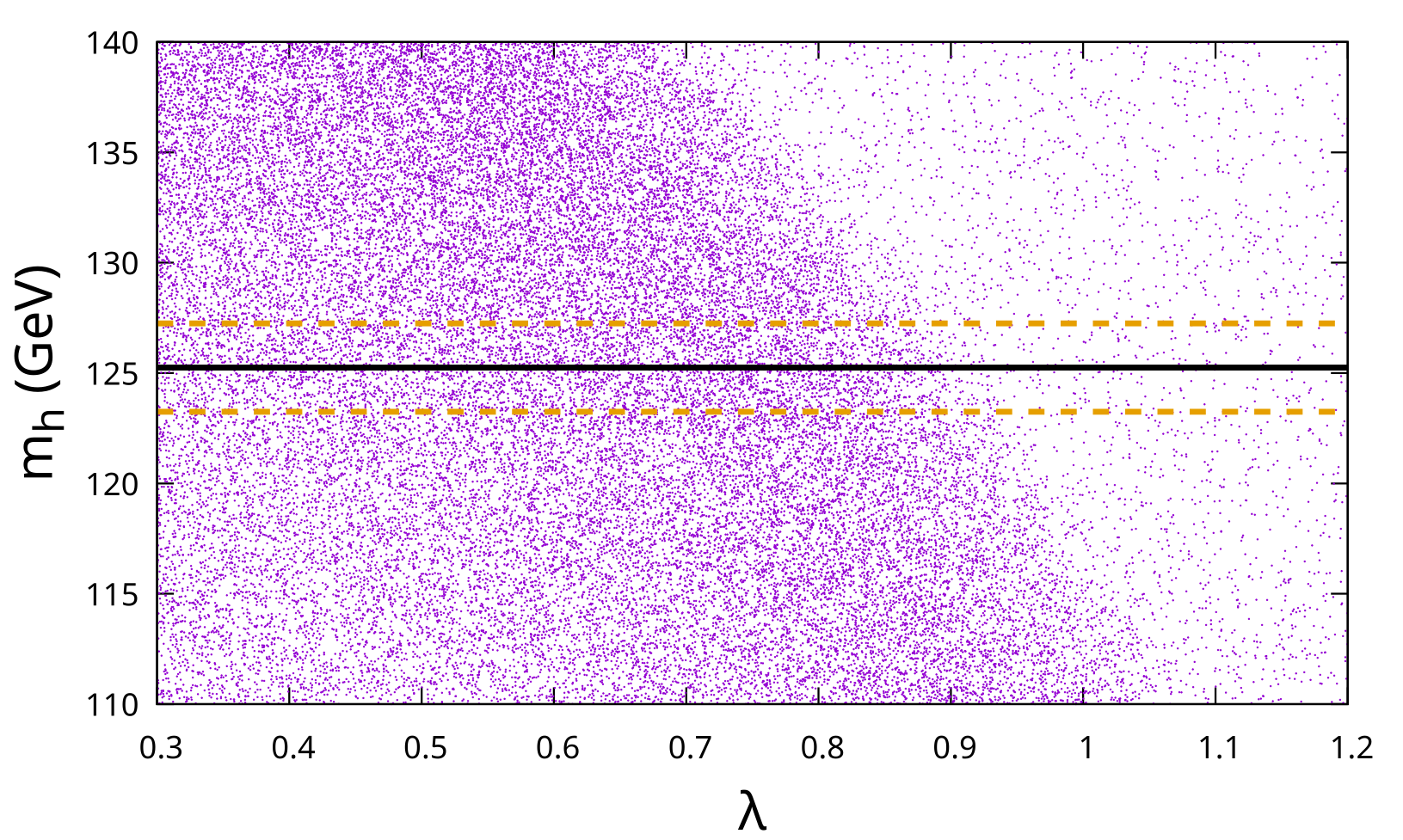}
%
% If no graphics program available, insert a blank space i.e. use
%\picplace{5cm}{2cm} % Give the correct figure height and width in cm
%
\caption{The light Higgs boson mass as a function of the trilinear parameter $\lambda$. The black line denotes the experimental value of the Higgs mass, $m_h=125.25$ GeV, while the orange dashed lines denote the 2~GeV theoretical uncertainties.}
\label{fig:higgsl}       % Give a unique label
\end{figure}

The light Higgs boson mass, $m_h$, as calculated for our model, is shown in \reffi{fig:higgsM} as a function of the unified gaugino mass $M_U$ and in \reffi{fig:higgsl} as a function of the trilinear coupling $\lambda$. Only points that agree with the experimental values of the top and bottom masses are included. We see that the unified~gaugino mass, $M_U$ cannot be more than $1800$GeV, and the~most points that satisfy the experimental limits on the Higgs~boson mass \cite{ParticleDataGroup:2022pth},
\begin{equation}
m_h^{exp}=(125.25\pm 0.17)~\text{GeV}\,, \label{higgsmass}
\end{equation}
are the ones that~have $1600\text{GeV}<M_U<1700$~GeV. 
 We also include~a theoretical uncertainty of 2~GeV
\cite{Slavich:2020zjv}.~Similarly, for the trilinear coupling $\lambda$ we see~that $\lambda<0.9$ is preferred in order to get the Higgs mass~values within the uncertainties.

\begin{figure}[htb!]
\sidecaption
% For example, with the graphicx style use
\includegraphics[scale=.5]{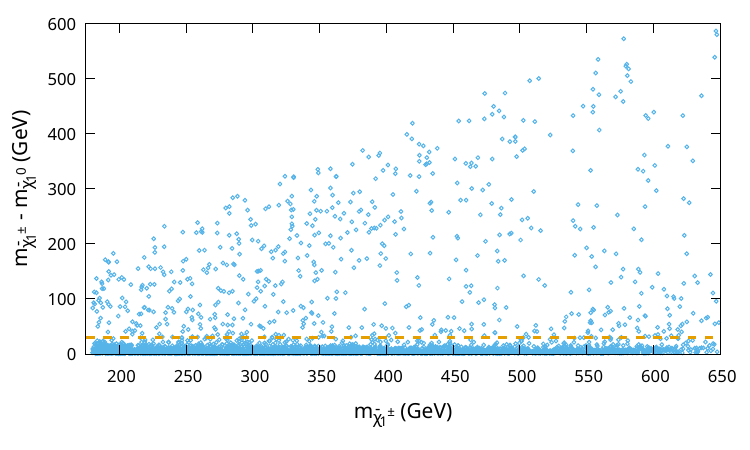}
%
% If no graphics program available, insert a blank space i.e. use
%\picplace{5cm}{2cm} % Give the correct figure height and width in cm
%
\caption{The plot  shows  the mass difference between the lightest chargino and the lightest neutralino, for points that satisfy the Higgs mass theoretical uncertainty of \cite{Slavich:2020zjv}. The orange dashed line denotes the 30~GeV mass difference limit.}
\label{fig:charneut}       % Give a unique label
\end{figure}

The difference between the lightest chargino mass and the
lightest neutralino, which  is the lowest supersymmetric particle (LSP) is given~in \reffi{fig:charneut} w.r.t. the
lightest chargino mass. For all points the Higgs mass is within the $2$ GeV theoretical uncertainty of \cite{Slavich:2020zjv} and satisfy the lower exclusion bounds for the  lightest chargino mass \cite{ParticleDataGroup:2022pth}. 
The ATLAS and CMS experiments of the LHC have searched for charginos and neutralinos and they have obtained bounds
of up to $1.4$ TeV, which, however, depend
on the mass difference among the lighter chargino and the lightest neutralino and, to some extent, also on
the details of the~decays 
\cite{ATLAS:2019lng,ATLAS:2021moa,CMS:2021edw,CMS:2021cox}.
The points below the~orange line
feature a chargino~mass of above 180 GeV 
and the mass~difference to the lightest
neutralino is~below 30 GeV, impyling
that these~points pass the experimental
bounds as~these are higgsino-like states. For the other points a more detailed~investigation is required which
we postpone to a future work.

\begin{figure}[htb!]
\sidecaption
% For example, with the graphicx style use
\includegraphics[scale=.5]{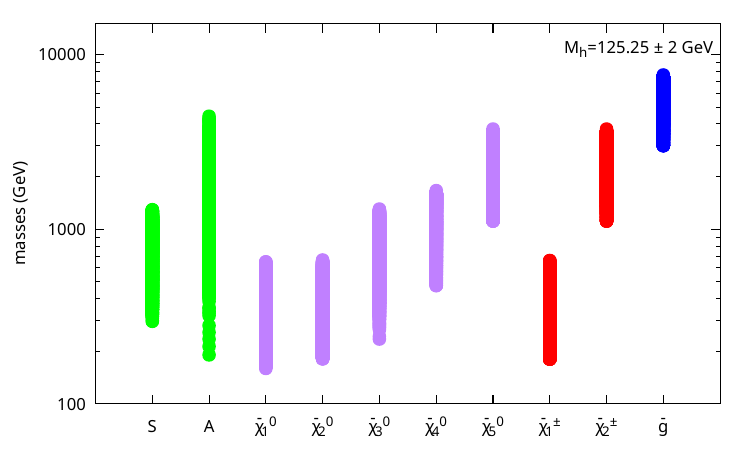}
%
% If no graphics program available, insert a blank space i.e. use
%\picplace{5cm}{2cm} % Give the correct figure height and width in cm
%
\caption{The plot shows  the predicted particle spectrum. The green points are the CP-even and CP-odd  singlet scalar masses; the purple points are the neutralino masses; the red ones are the chargino masses, followed by the blue points indicating the gluino masses.}
\label{fig:spectrum}       % Give a unique label
\end{figure}

 \reffi{fig:spectrum} features the predicted particle spectrum.
The (mainly) CP-even singlet scalar is denoted as $S$, while its CP-odd counterpart as $A$. Interestingly, in the parameter region in which our model agrees with the observed Higgs boson mass, $S$ is always heavier than  $\sim300$~GeV. The singlet component of the Higgs boson $h$ is sufficiently~small to be consistent with the existing coupling~measurements to vector bosons and fermions. 
They are not affected~by existing searches as they can hardly be produced at the LHC~because they are gauge singlets. 
$\tilde{\chi}_i^0, \tilde{\chi}_1^{\pm}$ and~$\tilde{g}$ are the neutralinos, charginos and the~gluinos, respectively. The points shown~correspond to the ones below the orange
line in~\reffi{fig:charneut} to ensure
that they~are compatible with existing
searches~at the LHC. Adding the other 
points~wouldn't change the picture significantly and the most important~change would be somewhat smaller values for 
mass of~the lightest neutralino.
Note, that~the gluino is predicted to be heavier than $2$ TeV in this model~and, thus, this model can explain why so far no sign for supersymmetry has~been found at the LHC. 
This also implies~that this model
will be difficult~to probe in the coming
LHC runs. The~reach of the high luminosity LHC for the lightest
chargino can~go up to 200 GeV if the
systematics~are well under control
\cite{Barducci:2015ffa}. Other 
possibilities~are the combined 
production of~a heavier neutralino 
together with~the lightest chargino 
which we will~investigate in an upcoming
work. 
Last but~not least we point
out that the~lightest neutralino is
an admixture~of the singlet fermion and a higgsino and thus it can be a cold~dark matter candidate consistent with
observations,~which we will investigate
together with~details of the collider searches.

%\begin{figure}
%\centering
%\includegraphics[width=0.8\textwidth]{phasma.pdf}
%\caption{\textit{The plot shows  the predicted spectrum for points with Higgs mass within the 2~GeV theoretical uncertainty and lightest chargino mass above 180~GeV. The green points are the CP-even and CP-odd  singlet scalar masses; the purple points are the neutralino masses; the red ones are the chargino masses, followed by the blue points indicating the gluino masses.}}
%\label{fig:spectrum}
%\end{figure} 

\section{Conclusions}\label{conclusions}

We reviewed a theory that starts from a $10D$, $\mathcal{N}=1$, $E_8$ Yang-Mills-Dirac theory constructed on the manifold $SU(3)/U(1)\times U(1)\times \mathbb{Z}_3$. The CSDR and Wilson  breaking mechanisms lead to a softly-broken $\mathcal{N}=1$, $SU(3)^3\times U(1)^2$, $4D$ effective  theory. The theory further breaks into a Split NMSSM scenario with top, bottom and light Higgs masses within the experimental limits. The model predicts that the gluinos are beyond the reach of the high-luminosity LHC. However, the lighter charginos and neutralinos are below the TeV in most cases with a small mass splitting. The reach of the high-luminosity LHC for scenarios with a somewhat larger mass splitting will be investigated in a future work.

\begin{acknowledgement}
GP is supported by the Portuguese Funda\c{c}\~{a}o para a Ci\^{e}ncia e Tecnologia (FCT) under Contracts UIDB/00777/2020, and UIDP/00777/2020, these projects are partially funded through POCTI (FEDER), COMPETE, QREN, and the EU. GP has a postdoctoral fellowship in the framework of UIDP/00777/2020 with reference BL154/2022\_IST\_ID.
GZ would like to thank the MPP-Munich and DFG Exzellenzcluster 2181:STRUCTURES of Heidelberg University for support.
\end{acknowledgement}
\end{document}